\documentclass[nofootinbib,twocolumn]{revtex4}
\usepackage{amsmath}
\usepackage{amssymb}
\usepackage{graphicx}
\usepackage[usenames, dvipsnames]{color}

\newcommand{\PT}{P\"oschl-Teller}
\newcommand{\xw}{x_{\rm wall}}
\newcommand{\phat}{\widehat{\Phi}}

\begin{document}

\title{Gravitational wave sources: reflections and echoes 
}

\author{Richard H.~Price} 
\affiliation{Department of Physics, MIT, 77 Massachusetts Ave., Cambridge, MA 02139}
\affiliation{Department of Physics, University
  of Massachusetts, Dartmouth, MA 02747}

\author{Gaurav Khanna} 
\affiliation{Department of Physics, University
  of Massachusetts, Dartmouth, MA 02747}

\begin{abstract}
The recent detection of gravitational waves has generated interest in
alternatives to the black hole interpretation of sources. A subset of such 
alternatives involves a prediction of gravitational wave ``echoes''. We consider 
two aspects of possible echoes: First, general features of echoes coming from 
spacetime reflecting conditions. We find that the detailed nature of such echoes 
does not bear any clear relationship to quasi-normal frequencies. Second, we 
point out the pitfalls in the analysis of local reflecting ``walls'' near the 
horizon of rapidly rotating black holes.
\end{abstract}

\maketitle

\section{Introduction}\label{sec:Intro}
The source of the recently detected gravitational waves (GWs) by the
LIGO collaboration~\cite{GWPRL1,GWPRL2} has been interpreted to be the
inspiral and merger of a pair of intermediate mass binary black
holes. This interpretation has been viewed as secure since the
observed waveform had an excellent fit to the very different physics
of early and late inspiral.  The early waveform fit the ``chirp''
pattern~\cite{chirp} of the evolving nearly circular binary orbit
driven to smaller radii and higher angular velocity by the loss of
energy to outgoing gravitational waves. The late waveform fit the
pattern for the quasinormal ringdown (QNR) of the perturbed final
black hole, a black hole of the appropriate angular momentum and mass 
as implied by the merger process~\cite{others}.

The early pattern is not exclusive to orbiting black holes; it would be no 
different for a binary of any compact objects of the same masses. What is most
important for the black hole interpretation is the QNR, and the way in which the 
transition from the early waveform to the QNR agrees with the black hole models
of numerical relativity~\cite{numrel}.

The importance of the QNR to the black hole interpretation has led to
the question of alternative, non-black hole, sources of QNR-like
waveforms~\cite{giddings,luciano,yunes,miller,CFandP}.  One recent
model of a source is the double wormhole  of Cardoso, Franzen and
Pani~\cite{CFandP} (hereafter CFP). The fact that damped oscillations
are not uniquely, or even especially, associated with black holes is
not news~\cite{nollert,kostas}, but a relatively new element of the
question is whether the replacement for the black hole may involve
reflections and may produce echoes, i.e., delayed repetitions of the
QNR-like pattern~\cite{CFandP,CHMPP,jren,BCG,SGMD}. Indeed, a discovery has
been claimed of just such echoes in the gravitational wave detector
data~\cite{echoesclaim,rerebuttal}, though the statistical
significance of the claim has been disputed by members of the LIGO
collaboration~\cite{rebuttal}.

In this paper, we do not focus exclusively on the LIGO detections, but
rather we consider somewhat broadly the physics that lies behind
recent claims, the nature of reflections of gravitational waves and
echoes that might result from such reflections from surfaces around
compact objects.  There is, however, a possible relevance to gravitational 
waveform interpretation: the issue of how closely echoes might be delayed
repetitions of an earlier burst.  Do echoes, for instance, have the
same frequency and damping rate of the late ``ringing'' in an initial
burst?  Might differences between the initial burst and its echo
contain, at least in principle, interesting information?

Clear answers to these questions may have strong implications 
for the claims made in Refs.~\cite{echoesclaim,rerebuttal} since that work 
relies on the echo signal sharing detailed characteristics with the initial 
burst from the black hole binary system. 

We discuss, in Sec.~\ref{sec:Nature}, the general nature of echoes,
and connect that issue to the meaning and features of quasinormal (QN) 
modes.  We shall point out the distinction between two very
different sources of echoes: On the one hand echoes can result from a
feature of the ``curvature potential'' through which waves
propagate~\cite{curvpot}; on the other hand echoes can be the result
of some sort of ``wall'' surrounding a compact object.

In Sec.~\ref{sec:Reflecs}, we pay particular attention to the physical
meaning of ``reflection,'' and point out a pitfall in the mathematical
analysis of refection of radiation at a surface around a black hole.
We conclude and summarize in Sec.~\ref{sec:Conc}.

Throughout, the paper we use the conventions of the textbook by Misner
{\it et al.}~\cite{MTW}.  In particular, we use the metric convention
-+++, and units in which $G=c=1$. For simplicity we will, for the most
part, use spherical symmetry in examples, so that, for instance, we
will give details for Schwarzschild, rather than the astrophysically
more relevant rotating Kerr holes. But issues of Kerr holes will be
important, and will constitute the motivation, especially in
Sec.~\ref{sec:Reflecs}.

%\pagebreak

\section{The nature of echoes from compact objects}\label{sec:Nature}
\subsection{Sources of echoes}\label{subsec:sources}
At the outset it is important to note that there can be at least two
distinct sources of echoes. One source is the spacetime itself, and
more specifically the curvature potential through which waves
propagate. An example of this is the double light ring model of
CFP~\cite{CFandP}. In that model, two peaks in
the curvature potential act, in effect, as two locations at which wave
interactions can be viewed in terms of transmission and reflection. A
second source of echoes is some sort of a ``wall'' that forms an
inner boundary of the wave propagation problem, and that replaces the
horizon as the boundary~\cite{echoesclaim,otherwalls}. These walls are
typically associated with speculations, or specific models, of quantum
effects.

It is crucial to emphasize here the difference between formal
quasinormal ringing (QNR) and quasinormal-like oscillations
(QNR-like). The former refers to an eigenvalue problem for single
frequency modes of a system, typically a system characterized by a
fairly compact potential for wave propagation. The boundary conditions
on the modes involve outgoing radiation, so that the eigenproblem is
not self adjoint, and the frequency eigenvalues are complex. In the
case of black holes, the boundary conditions are outgoing radiation at
infinity and ingoing radiation at the horizon. These complex
eigenvalues also show up as poles in the frequency-domain Green
function for the system. Typically, there is an infinite spectrum of
such modes for any linear system, e.g., for the differential equation
for a particular multipole mode of a black hole perturbation
field~\cite{KerrDifferent}.

The QNR-like signals are damped oscillations.  In the black hole
context these QNR-like waveforms have long been associated with the
late-time ``ringdown'' of perturbed black holes. While this
association developed in  work on black hole perturbation theory in
the 1970s, this ringdown has been seen in all numerical relativity
simulations of black hole ringdown, simulations based on the fully
nonlinear equations of Einstein's general relativity. Such QNR-like waveforms
typically have very nearly the period of oscillation and the exponential damping 
rate of the least damped of the quasinormal modes, and there was little
attention given to the difference.

In fact, a system with a time dependent source cannot exhibit pure
QNR~\cite{norinitialdata}. The outgoing signal will always be affected by the time
dependence of the source as well as the damped-sinusoid pattern of a
QN mode. From the Green's function point of view, the integral of the
source over the Green's function~\cite{GFrefs} will include a residue
for the QN pole, but will have other contributions.  It must be asked,
then, why is there such a close apparent correspondence between the
late time signal and the least damped QN mode? 

Part of the answer is that the correspondence is not always
valid. Nollert~\cite{nollert} studied the mathematical problem of
evolving initial data in the Schwarzschild spacetime and showed that a
class of minor modifications of the problem had no discernible effect
on the evolved data, but changed the QN spectrum enormously.  More
recently, CFP have shown that the
QN spectrum of a wormhole consisting of two Schwarzschild ``funnels''
is enormously different from that of the Schwarzschild black hole, yet
the initial QNR-like ringing of the wormhole is almost identical to
that of the black hole.

There are, therefore, examples in which there are weak or missing
connections between the QN frequencies of a system and the QNR-like
ringing exhibited in signals generated by sources. But there are
examples in which there is a strong connection and black hole
processes fall in that second class. It is important to ask why.

CFP have ascribed the QN frequency and QNR-like
ringing to the role of the light-ring. In the case of black holes this
is an interesting heuristic insight and one that was first shown to
give good estimates of QN frequencies by Goebels~\cite{Goebels}. It
cannot, however, be the complete story. One can, after all, trivially set 
up a 1+1 model (one spatial dimension, one time dimension) with outgoing
radiation boundary conditions but with no attached concept of a light
ring; such a problem will have a QN spectrum and a QNR-like
ringing. We have also presented a 3+1 model with no light ring, yet
with QNR-like oscillations~\cite{maybeourpaper}.

A more general view of the connection between QN and QNR-like
mathematics is that the QNR-like signal is due to ``scattering''
within a potential. That scattering can account for the damping of the
outgoing radiation.  The scattering viewpoint is very insensitive to
distant boundary conditions and should be initiated as the source
(initial data or particle motion) interacts with a peak of a
potential.

The scattering viewpoint suggests that a WKB approximation may give
good estimates, but of the frequencies of the QNR-like oscillations,
{\em not} of the true QN eigenvales.  The WKB approximation uses an
integral over the potential, so it is insensitive to the changes
(e.g., those of Nollert) that greatly change the QN spectrum; the
approximation also is insensitive to distant boundary conditions.  To
some extent the WKB approximation and the scattering viewpoint are
conceptually, or heuristically quite close. This can be taken as a
partial explanation of the examples in which the QN frequencies do not
agree with the scattering/WKB results. This disagreement is most
pronounced when the curvature potential is not smoothly varying in
space. The condition for success of the WKB approximation is that the
spatial rate of change of the curvature potential is small~\cite{MW}.

The WKB approximation has given fairly good agreement with computed
black hole QN frequencies, but it must be kept in mind that the WKB
approximation is a high frequency approximation, and it is typically
applied to wavelengths that are of order of the width of the potential
that affects wave propagation. The situation then is that we can take
some comfort in the WKB approximation giving results in good agreement
with computed waveforms, but must not be surprised in the absence of
such agreement.

This scattering viewpoint lets us make some predictions about the
nature of the QNR-like signals in echoes. These will be discussed
below.

\subsection{A model problem: the P\"oschl-Teller potential with a reflecting 
wall}\label{subsec:PT}
The work by CFP has provided useful examples of echoes from a
potential with two peaks. This is one of two distinct ways in which
echoes can be generated. We will refer to that paper in arguments
below, but here we shall focus on the other general manner in which
echoes can be generated: a reflecting wall. Our specific model will
start with the equation
\begin{equation}\label{genDE}
 \frac{\partial^2\Psi}{\partial t^2}
 - \frac{\partial^2\Psi}{\partial {x}^2}+V(x)\Psi={\rm Source}\,.
\end{equation}
For a Schwarzschild black hole $\Psi$ is a representation of a
multipole of a scalar, electromagnetic or gravitational perturbation
field; the $x$ coordinate is the Regge-Wheeler~\cite{RW57} tortoise
coordinate $r^*$, and the source term can represtent a particle.  The
potential, in the black hole case is the curvature
potential~\cite{curvpot}, which falls off as $1/{r^*}^2$ as
${r^*}\rightarrow\infty$, and falls off exponentially as
${r^*}\rightarrow-\infty$, the location of the horizon.

For our model we will start with the P\"oschl-Teller
potential
\begin{equation}\label{eq:PT}
  V_{PT}(x)=1/\cosh^2{x}\,.
\end{equation}
We  will put this well studied model~\cite{PT,ferrarimashhoon} to a new purpose
by imposing nonstandard boundary conditions.
As  $x\rightarrow\infty$, we use the standard outgoing condition
%The typical boundary conditions are outgoing condition at
%$x\rightarrow\infty$, 
i.e., $\Psi$ becomes proportional to
$\exp{[i\omega(x-t)]}$.  For the other
boundary condition, the standard choice is $\Psi\propto\exp{[i\omega(-x-t)]}$ as
$x\rightarrow-\infty$, which we will call the horizon condition, since
it is the analog of the horizon boundary condition for black
holes. But we may also take, as a model for reflection, a ``wall
condition,'' the condition that $\Psi=0$ at some particular value of
$x$, the location of a reflecting wall. 

The solution of the system of Eqs.~(\ref{genDE}) and (\ref{eq:PT}),
with the outgoing condition, is proportional to the associated Legendre
function
\begin{equation}
  \Psi\propto P^\nu_\mu(\tanh{x})\quad\quad \nu=\frac{1}{2}(-1\pm
  i\,\sqrt{3}) \quad\quad \mu=i\omega\,.
\end{equation}
By expressing this in terms of a Gauss hypergeometric function it can
be shown that the horizon condition at $x\rightarrow-\infty$ is
achieved only for
\begin{equation}\label{eq:QNPTomegas}
  \omega=-i\left(n+\frac{1}{2}\right)\pm\frac{\sqrt{3\;}}{2}\,,
\end{equation}
where $n=0,1,2,...$. The frequency of interest, the least damped QN
mode, is that for $n=0$.

In the case of the reflecting wall condition, $\Psi=0$ at $x_{\rm
  wall}$, we must search numerically for the complex value of $\omega$
for which $P^{(-1\pm i\sqrt{3\;})/2}_{i\omega}(\tanh{x_{\rm
    wall}})=0$. This search was carried out by a simple code that
performs a series of hierarchical searches over the complex plane
looking for zeros of $P^{(-1\pm i\sqrt{3\;})/2}_{i\omega}(\tanh{x_{\rm
    wall}})$ as needed. We begin with a coarsely refined grid, and
identify regions of the plane that yield values of the mentioned
function that are lower than a certain threshold. We then use these as
the center points for more refined grids and resume the search. This
process is continued several times (the search threshold value is
lowered with each refinement, of course). This allows us to home in on
the zeros of the function in fairly simple and accurate manner. A
separate numerical exercise was to evolve $\Psi$, from initial data
representing a narrow Gaussian pulse, starting at  $x=25$ (large enough so that
the potential is effectively zero) and
moving in the negative $x$ direction.  This was done using a separate
time-domain wave-equation solver that uses a time-explicit, 2-step Lax
Wendroff, second-order finite-difference evolution scheme.

%FIT FOR HORIZON -.0095*cos(.866*(x-85.2))*exp(-0.5*(x-84.57))

\begin{figure}[h]
  \begin{center}
  \includegraphics[width=.45\textwidth ]{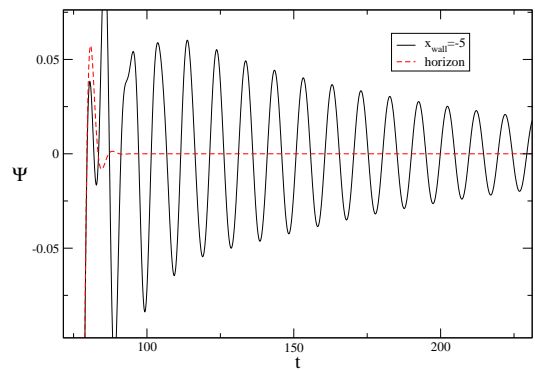}
  \caption{The late time waveform evolved from a narrow inward moving
    initial Gaussian pulse. The dashed curve is the waveform for the
    horizon condition (i.e., the extended P\"oschl-Teller potential);
    the solid curve is for the P\"oschl-Teller potential plus a wall
($\Psi=0$) condition at $x=-5$.}
  \label{fig:comparo5vshor}
  \end{center}
  \end{figure}

Our first example appears in Fig.~\ref{fig:comparo5vshor}. The
evolutions of an initial ingoing Gaussian pulse are shown for both the
pure P\"oschl-Teller potential (dashed curve), and for the
P\"oschl-Teller potential truncated by a reflecting wall at $x=-5$
(solid curve).
Because the reflecting wall is so close to the peak, this model does
not involve echoes, but rather changes the problem to another with a
modified single peak.  This example serves to show a case in which the
QNR-like ringing agrees quite well with the QN eigenvalue/pole. The
pure P\"oschl-Teller potential has, according to
Eq.~(\ref{eq:QNPTomegas}), a least damped QN $\omega$ of
$(\pm\sqrt{3\;}-i)/2$, which agrees to good accuracy
with a fit to the dashed curve in the Fig.~\ref{fig:comparo5vshor}
for $x$ larger than around 80.  The eigenvalue search for the
$x_{\rm wall}=-5$ case gives a least damped value of 0.640 + i 0.0096,
which agrees with a fit to the curve to about 1\% with the real part
and to about 5\% (the fit uncertainty) for the imaginary part.

It will be of some interest, for later models, to check the applicability 
of the Schutz-Will WKB approximation~\cite{schutzwill} for the dominant 
(i.e., least damped) QN frequency $\omega_{\rm QN}$:
\begin{equation}\label{SchWill}
  \omega^2_{\rm QN}=
V-i\,\frac{1}{2}\sqrt{-2\frac{d^2V}{dx^2}\;}\,.
\end{equation}
This formula is to be applied at the peak of the potential, where the
second derivative of the curvature potential is negative, and hence
the second term on the right is pure imaginary. In the case of the
\PT\ potential in Eq.~\eqref{eq:PT} this gives 
$\omega_{\rm  QN}=1.0987-i\,0.4551$. Note that this result is a reasonable
approximation of the pure \PT\ QN mode. 

The WKB prediction $\omega_{\rm QN}=1.0987-i\,0.4551$ applies to the
model with $x_{\rm wall}=-5$ as well as to the pure \PT\ potential,
since both models in Fig.~\ref{fig:comparo5vshor} have the same peak
behavior. Here the WKB approximation is still in the right ballpark for
the real part, but orders of magnitude wrong for the imaginary
part. This should be expected. The Schutz-Will estimate approximates the 
effective curvature potential as a parabola near the peak and works best
if the turning points, those locations at which $\omega^2=V(x)$, are close 
together. For the high QN frequencies in some models, this does not apply; there
are not even any turning points. It is not surprising that the real part 
in the Schutz-Will estimate Eq.~\eqref{SchWill}, which does not depend
delicately on the shape of the potential is widely applicable, though it 
is surprising how good an estimate it is.

It is worth emphasizing that the WKB method is local; it may be considered
to be related to the scattering picture of  QNR-like phenomena. It may also
be worth emphasizing that in our \PT\ model there is no meaning to a ``light ring.''

\begin{figure}[h]
  \begin{center}
  \includegraphics[width=.45\textwidth ]{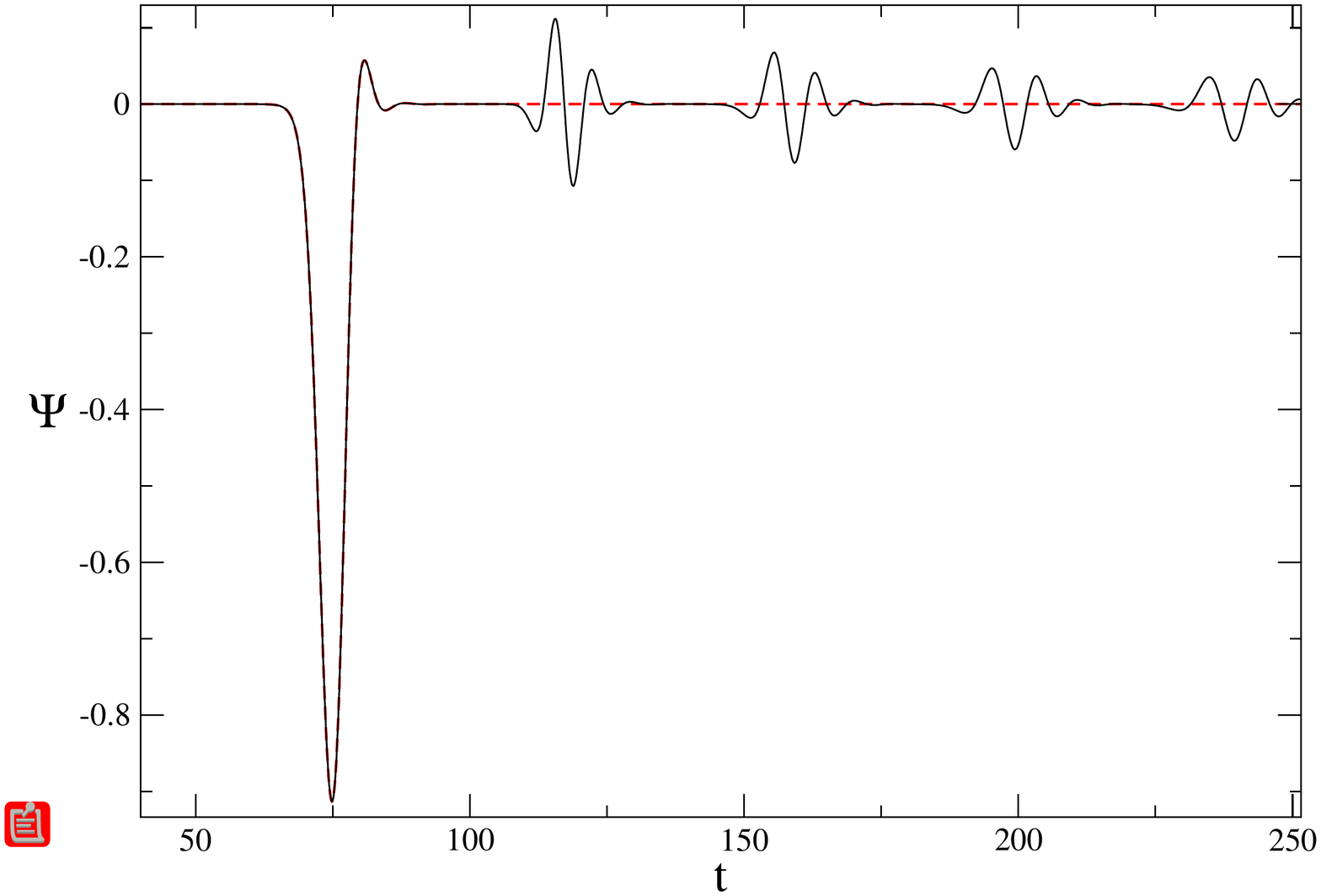}
  \caption{The waveform from an initial narrow Gaussian pulse evolved
    both with a pure \PT\ potential (dashed curve), and with a reflecting wall at
    $x_{\rm wall}=-20$ (solid curve).  }
  \label{fig:comparo_hor_vsm20}
  \end{center}
\end{figure}

Results are shown in Fig.~\ref{fig:comparo_hor_vsm20} comparing the
waveform evolved from an initial Gaussian pulse for both the pure
\PT\ potential, and the \PT\ potential with a reflecting wall at
$\xw=-20$. In the reflecting wall case there is an initial burst that
is essentially indistinguishable in the graph from the burst evolved
with the pure \PT\ potential. This is a particularly clear example of
the distinction between a QN oscillation and a QNR-like oscillation. There 
are multiple QN modes in this model. The one that appears to most relevant 
for the model with $\xw=-20$ is $0.936 + i\,0.01$.\footnote{Note the small 
imaginary part; the other modes have an imaginary part that is even smaller.
We enlist a few additional modes here: $0.784 + i\,0.0056$, $0.631 + i\,0.0024$, 
$0.476 + i\,0.0009$.} The QNR-like first burst, however, accurately traces
the pure \PT\ burst, which has both a QN frequency, and an evolved wave
form with $\omega=0.866+i\,0.5$. Again, we see that the scattering viewpoint
is justified, and the real part of the WKB approximation is correct to rough 
order. 

The question remains on the nature of the echoes in the reflecting wall 
case. It might be expected that later and later echoes would approach more
and more closely the true QN frequency. In Fig.~\ref{fig:comparo_hor_vsm20}, 
however, there is no sign that the echoes approach the almost undamped oscillations 
of the true QN mode.

\begin{figure}[h]
  \begin{center}
  \includegraphics[width=.45\textwidth ]{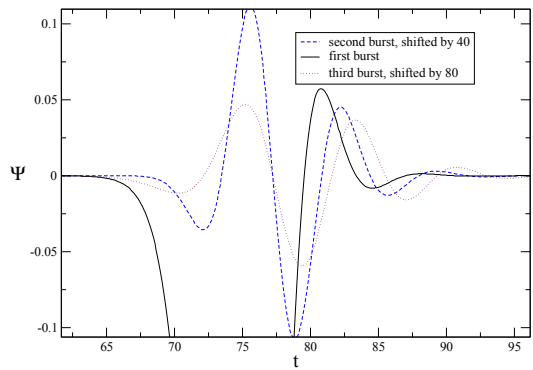}
  \caption{ The waveform from an initial narrow Gaussian pulse evolved
    both with a reflecting wall at $x_{\rm wall}=-20$. The first
    echo is shifted by 40 and the second echo by 80, to bring those
    echoes in approximate alignment with the first burst. }
  \label{fig:m20shfited}
  \end{center}
\end{figure}

The relationship of the echoes and the first burst is examined in
Fig.~\ref{fig:m20shfited}. If we consider outgoing radiation to be
generated at $x=0$, and the first echo to be the reflection off the
wall at $\xw=-20$, then the outgoing echo should follow the initial
QNR-like burst by a time delay of 40. For that reason, in
Fig.~\ref{fig:m20shfited} we shift the first echo to an earlier time
by 40. For the same reason we shift the second echo to earlier time by
80. The curves in Fig.~\ref{fig:m20shfited} show that the basic idea
of a delayed echo is correct, but that the delay time is somewhat
larger than 40 for each ``bounce.''

The exponential damping rate of the first echo is neither the 0.5 of
the pure \PT\ QN, nor the 0.01 of the QN for the reflecting
potential, but rather a value around 0.35. The appropriate $\omega$
for the QNR-like echo cannot be extracted with good precision because
the late-time portion of the echo is not well approximated by a single
damped sinusoid. We have made arguments
above that a pure QN oscillation is impossible, since the source has
its own time variation. {\em We conjecture that the echoes, QNR-like
ringing present in outgoing radiation, are not pure QN oscillations}.
Independent of that conjecture is the fact that in principle there is
a difference between the shape of the initial burst and that of any of
the echoes. If we generalize from this one example, we can conclude
that {\em the echo waveforms contain important information about the
conditions from which the echoes emerge.}

\begin{figure}[h]
  \begin{center}
  \includegraphics[width=.4\textwidth ]{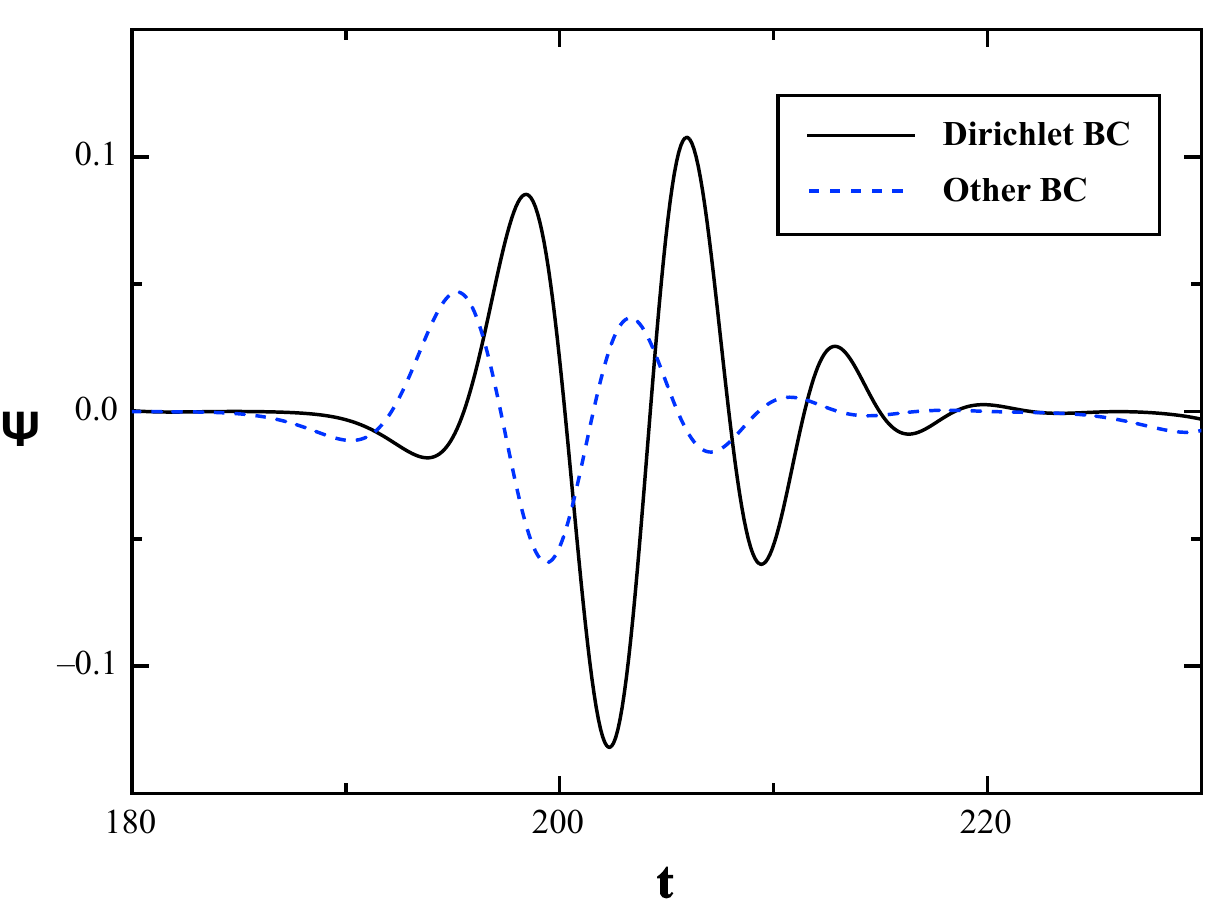}
  \caption{ The third echo from an initial narrow Gaussian pulse evolved
    with a reflecting wall (solid curve), and with a different boundary 
    condition (dashed curve) at $x_{\rm wall}=-20$. It is clear that 
    they exhibit very different characteristics.}
  \label{fig:compareBCs}
  \end{center}
\end{figure}

To show more evidence in support of our conjecture, we changed the 
boundary condition from a reflecting wall, i.e. a Dirichlet boundary, 
to a different condition -- one that effectively relates the second  
time-derivative of the field to the negative of the field itself. The 
results are shown in Fig.~\ref{fig:compareBCs}. 
It is clear that the echoes with this boundary condition are very different 
from those from the reflecting wall. Therefore, these echoes carry 
important detailed information about the processes that led to their 
formation and development. Very recently, i.e. well after this current work 
was completed, a new preprint from Mark et al.~\cite{mark} appeared that 
performs an extensive study of echo waveforms in the context of a Schwarzschild 
hole via a Green's function approach. While the goal of that effort is to 
provide a template for the echo signals, they confirm several of the claims 
and conjectures that we make in this work.   

Now, since one is most interested in considering echoes in the context of
gravitational waves (as opposed to the scalar case, considered thus far) 
from a realistic, spinning black hole system, one may attempt to simply go 
ahead and apply a Dirichlet-type boundary condition at a location close to 
the horizon on the Weyl scalar $\Psi_4$ and evolve using the Teukolsky 
equation~\cite{teuk}. However, such a naive attempt yields no echoes whatsoever! 
In the next section, we explain this null result and also sketch out a potential 
scheme for attempting to implement a proper reflection condition in the context 
of gravitational waves. \\

\section{Reflections of gravitational waves}\label{sec:Reflecs}
In this section we consider the description of reflections at some
sort of ``wall.'' We shall not be, nor need to be specific about the
nature of this reflecting wall, except for one requirement. The
reflection must be the result of a local condition, and not a
condition like the modification of the potential (that is in a loose
sense, global).  We shall clarify what we mean by this with examples
of electromagnetism and gravitational perturbations in the
Schwarzschild background.

The fundamental concept we want to present here is that, except for a
scalar field, there are many features of a field that are encoded in
different mathematical packages. Because a gravitational perturbation,
with its 10 degrees of freedom is an unnecessarily complicated way to
start, but a scalar perturbation is too simple, the complexity
``Goldilocks zone'' is occupied by electromagnetic perturbation of a
Schwarzschild background.

In this case, at any point in spacetime, there are 6 degrees of
freedom that can be considered to be the 3 components of the electric
field, and the 3 of the magnetic field; alternatively they can be
considered the 6 independent components of the Maxwell 4-tensor.

The partial differential equations for this system, Maxwell's
equations in the Schwarzschild background, are uselessly messy in
terms of the individual vector or tensor components. A very effective
way of repackaging these quantities is to use the 3 complex fields of
the Newman-Penrose (NP) formalism.  The asymptotic behaviors of these
fields, and hence the argument to be made here, depend crucially on
the spin-weight of the fields. For that reason we use here a notation
that indexes the fields with their spin-weight~\cite{PriceII}. The
formal definitions of these complex fields, and their connection to
the original NP notation, are given in Appendix A in that reference.

The 3 complex fields can most simply be defined through their
relationship to the components on the electric and magnetic fields in
the Schwarzschild background. We define
$E^{[r]},E^{[\theta]},E^{[\phi]}$ as the orthonormal components of the
electric field in the basis given by the standard Schwarzschild
($r,\theta,\phi$) coordinate system. The components of the magnetic
field are similarly defined with a $B$. The NP projections
$\Phi_{-1},\Phi_{-1},\Phi_{+1}$ are related to these $E,B$ components
by

\begin{widetext}
 \begin{eqnarray}
{\Phi}_{+1}&=&2^{-1/2}\left(1-2M/r\right)^{-1/2}\left[
  \left(E^{[\theta]}-B^{[\phi]}\right)
  +i\,\left(E^{[\phi]}+E^{[\theta]}\right)\right]\label{PhEBa}\\ 
{\Phi}_{0}&=&-\frac{1}{2}\left(E^{[r]}+i\,B^{[r]}\right)\label{PhEBb}\\ 
{\Phi}_{-1}&=&-2^{-3/2}\left(1-2M/r\right)^{1/2}\left[
  \left(E^{[\theta]}+B^{[\phi]}\right)
  -i\,\left(E^{[\phi]}-E^{[\theta]}\right)\right].\label{PhEBc}
\end{eqnarray}
\end{widetext}

These relations point to an important property of the $\Phi_{k}$:
their relationship to ingoing and outgoing radiation. Consider, for
example, the quantities constructed on the right in
Eq.~(\ref{PhEBa}). For outgoing electromagnetic radiation, the
orthonormal components of the electric and magnetic fields all fall
off as $1/r$, but $E^{[\theta]}=B^{[\phi]}$ and
$E^{[\phi]}=-B^{[\theta]}$ to leading order in in $1/r$, so that to
this order $\Phi_{+1}$ vanishes. It turns out, in fact, that
$\Phi_{+1}$ falls off in the large $r$ limit as $1/r^3$. More
generally, there is a ``peeling theorem'' for the $\Phi_{k}$ that
tells us that~\cite{penrosepeeling}
\begin{equation}\label{largerlim}
\Phi_k\xrightarrow{r\rightarrow\infty}   1/r^{2+k}\,.
%\stackrel{r\rightarrow\infty\ \ \ }{\Phi_k\rightarrow 1/r^{2+k}}\,.
\end{equation}
It is then $\Phi_{-1}$ that describes outgoing radiation. In that
sense it plays the role of $\psi_4$ in the Teukolsky
equation~\cite{teuk}, the quantity that describes outgoing radiation.

In the same sense, there is a version of a peeling theorem in the
horizon limit.  It can be shown~\cite{invpeel} that in the horizon
limit, i.e., in the limit $r^*\rightarrow-\infty$,
\begin{equation}
\Phi_k\rightarrow \exp{(-kr^*/2M)}\,.
\end{equation}
The quantity that is dominant in the description of radiation being
carried into the horizon is therefore $\Phi_{+1}$.

Although $\Phi_{-1}$ is dominant for outgoing radiation, and
$\Phi_{+1}$ for ingoing, each of the $\Phi_k$ carries all information
about the other $\Phi_k$. To express these relationships it is best to
consider individual multipoles and remove the angular dependence. The
angular dependence of NP fields is described with spin-weighted
spherical harmonics. We denote by a caret (\;${\widehat{\ } }$\;) the
function of $r,t$ multiplying each spin-weighted spherical
harmonic. (For the precise procedure for moving angular dependence,
see Ref.~\cite{PriceII}.)

These equations, i.e. the Maxwell differential equations, are best 
expressed in derivatives with respect to retarded and advanced time,
\begin{equation}
u=t-r^*\quad\quad v=t+r^*\
 \end{equation}
and in Gaussian-esu units.
For an $\ell$-pole mode the equations are 

\begin{widetext}
\begin{eqnarray}
2(1-2M/r)^{-1}\partial_v\left(r\phat_{-1}\right)&=&-\frac{1}{2}
\ell(\ell+1)\phat_{0}\label{Max1}\\
2(1-2M/r)^{-1}\partial_v\left(r^2\phat_{0}\right)&=&r\phat_{+1}\label{Max2}\\
\partial_u\left(r^2\phat_{0}\right)&=&r\phat_{-1}\label{Max3}\\
\partial_u\left[\left(1-2M/r\right)r\phat_{+1}\right]&=&
-\frac{1}{2}\ell(\ell+1)(1-2M/r)\phat_{0}\,.\label{Max4}
\end{eqnarray}
\end{widetext}

From these equations, second-order wave equations can be formulated
for any of the $\phat_k$, and all information could be extracted from
that $\phat_k$. We could then, in principle, work only with
$\phat_{+1}$ for outgoing radiation. By differentiating with respect to
$u$ we could find $\phat_0$, and then with a second differentiation
with respect to $u$ we could find $\phat_{-1}$.

This nature of the NP formalism, this separation into ingoing and
outgoing quantities is crucial to implementing reflection conditions.
The example of electromagnetic waves is instructive. The condition on
a perfectly conducting surface is the vanishing of the tangential
electric component and the normal magnetic component. 

For definiteness, let us consider even parity fields; these turn out
to involve only the real part of the $\Phi_k$ quantities. The
condition that the locally measured value of $E^{[\theta]}$ vanish,
requires both $\phat_{+1}$ and $\phat_{-1}$. For a reflecting surface
close to the horizon, i.e., at a large negative value of $r^*$, This
is numerically awkward since in the horizon limit the $\phat_{+1}$
diverges, and $\phat_{-1}$ vanishes.  If, for example, we use a wave
equation for $\phat_{-1}$, the boundary condition would require,
according to the Maxwell equations, \eqref{Max1} -- \eqref{Max4}, both
$\phat_{-1}$ and its second derivative with respect to advanced time.
Notice too that Eq.~\eqref{largerlim} tells us that a similar
awkwardness applies to a reflecting surface at large $r$.

Heuristically, this awkwardness can be traced to the fact that the
reflection condition involves a balance of ingoing and outgoing
radiation, and the NP quantities are specific to one or the other.
This suggests that reflection problems are best handled in a
computation using the ``balanced'' NP field $\phat_{0}$. From
Eqs.~\eqref{PhEBa} -- \eqref{PhEBc} and \eqref{Max1} --
\eqref{Max4}, it then follows that the no-reflection boundary
condition is simply that the derivatives of $r\phat_{0}$ with respect
to advanced and retarded time are opposites of each other.

We now turn to the problem of reflection of gravitational waves. 
The analog of the electromagnetic reflection conditions would be some conditions
on the transverse traceless components of gravitational strain. We need not 
know precisely what the reflection condition is, only that it is some {\em local} 
condition on the gravitational strain. 

The NP formalism for gravitational perturbations~\cite{chandra}
encodes all the information about the Weyl tensor in 5 complex fields,
$\Psi_{4},\Psi_{3},\Psi_{2},\Psi_{1},\Psi_{0}$, with properties
analogous to the 3 complex electromagnetic fields. In particular, $\Psi_{4}$ describes
outgoing radiation (the other $\Psi_k$ fall off faster than $1/r$ as
$r\rightarrow\infty$). Similarly, $\Psi_{0}$ describes ingoing
radiation, and the other $\Psi_k$ fall off faster than $\Psi_{0}$ as
$r^*\rightarrow-\infty$.

There is an important difference between the NP formalism for
gravitational perturbations of the Schwarzchild spacetime and those of
the NP formalism for electromagnetic perturbations. For
electromagnetism, all the NP projections of the Maxwell tensor are
gauge invariant; for gravitational perturbations, only $\Psi_4$ and
$\Psi_0$ are gauge invariant. The other $\Psi_k$ change under a
perturbative transformation of coordinates or projection tetrads. This
is why only $\Psi_4$ and $\Psi_0$ can be uncoupled from the other
$\Psi_k$ and made to satisfy single-unknown wave equations.

A wave equation for $\Psi_4$, in the context of Schwarzschild spacetime, 
uncoupled from the other $\Psi_k$ is known as the Bardeen-Press 
equation~\cite{BardeenPress}.  A physically motivated reflection 
condition near the horizon will involve both $\Psi_4$ and $\Psi_0$ 
in a manner analogous to the electromagnetic condition involving 
$\phat_{-1}$ and $\phat_{+1}$. One possibility for dealing with the 
local boundary conditions is for example, solve for $\Psi_4$, 
and from the solution find $\Psi_0$. In electromagnetism,
finding $\phat_{+1}$ from $\phat_{-1}$ required two derivatives with
respect to advanced time, and was numerically delicate. For
gravitational perturbations the situation is worse; finding $\Psi_0$
from $\Psi_4$ requires four differentiations with respect to advanced
time.

For gravitational perturbations of the Schwarzschild spacetime with
reflection conditions, the difficulty can be avoided by using the
Zerilli or Regge-Wheeler equations, which, like $\phat_0$ in the
electromagnetic case, are not skewed to ingoing or outgoing wave
propagation. Rapidly rotating black holes, however, do not provide
this easy workaround. For gravitational perturbations of the Kerr
spacetime, there exist no wave equations analogous to the
Regge-Wheeler or Zerilli equations; equations exist only for the gauge
invariants $\Psi_4$ and $\Psi_0$. Thus, for studies of reflections
from exotic ``walls'' near the horizon, either a very difficult
numerical boundary condition can be implemented, or it can be assumed
that the the results for the Schwarzschild background give adequate
insight for rapidly rotating holes. The former is much easier to implement 
in the frequency-domain, as attempted in Ref.~\cite{NSTT}. 

Another challenge associated to a study of echoes in rotating spacetimes via 
the Teukolsky equation arises from the lack of Birkhoff's theorem there.  
Does the Teukolsky equation even represent the evolution of perturbations 
of a compact object other than a Kerr black hole? This important consideration 
was recently raised by the authors of Ref.~\cite{mark}.

\section{Conclusions}\label{sec:Conc}
In this article we sought to clarify two aspects of 
``echoes'' in gravitational wave signals from the late stages of
binary inspiral.

The first is the general nature of echoes and their relationship to QN
modes. We point out that in a sequence of echoes, later echoes are not
copies of the first burst. This potentially has strong implications for the 
claims made in Refs.~\cite{echoesclaim,rerebuttal} since that work relies 
on the echo signal being a repetition of the initial burst. Furthermore, 
later and later echoes of an infinite string of echoes, do not approach a 
ringing at a QN frequency.  In general, a scattering viewpoint involving the 
curvature potential, where applicable, gives a better heuristic view of the 
process of signal generation than a QN analysis or considerations of a light 
ring.

The second goal of this paper is to warn of a pitfall in using the
Teukolsky~\cite{teuk} wave function $\Psi_4$ for analyzing the effect
of reflecting ``walls'' outside the horizon. Simply setting Dirichlet
or Neumann conditions on this wave function, for example, is not an 
expression of a locally reflecting wall. Moreover, as pointed out by 
Ref.~\cite{mark} the lack of Birkhoff's theorem in the context of a 
rotating spacetimes poses serious concerns on whether or not $\Psi_4$ 
is even the relevant quantity to study. 

{\em Acknowledgments:} We thank Alessandra Buonanno for suggesting that we 
explore these issues in detail. We also thank Kostas Kokkotas and Sebastian 
Voelkel for making helpful comments and suggestions that helped improve this
paper. G.K. acknowledges research support from NSF 
Grants No. PHY-1606333 and No. PHY-1414440, and from the U.S. Air Force 
agreement No. 10-RI-CRADA-09. 

%\pagebreak


\begin{thebibliography}{11}
\bibitem{KerrDifferent} For the Kerr spacetime, which does not have
  multipole modes in the usual sense, the details of the eigenvalue
  problem are somewhat different.
\bibitem{norinitialdata} QNR can also be evident in the evolution of
  initial data, but in this case also pure QNR is impossible. The initial data
for a pure QN mode is unbounded at infinity and at the horizon.
\end{thebibliography}
\end{document}